# Sensitivity of synthetic aperture laser optical feedback imaging

Wilfried Glastre,* Eric Lacot, Olivier Jacquin, Olivier Hugon and Hugues Guillet de Chatellus

*Centre National de la Recherche Scientifique / Université de Grenoble 1,*

*Laboratoire Interdisciplinaire de Physique, UMR 5588,*

*Grenoble, F- 38041, France*

*Corresponding author: wilfried.glastre@ujf-grenoble.fr

In this paper we compare the sensitivity of two imaging configurations both based on Laser Optical Feedback Imaging (LOFI). The first one is direct imaging, which uses conventional optical focalisation on target and the second one is made by Synthetic Aperture (SA) Laser, which uses numerical focalisation. We show that SA configuration allows to obtain good resolutions with high working distance and that the drawback of SA imagery is that it has a worse photometric balance in comparison to conventional microscope. This drawback is partially compensated by the important sensitivity of LOFI [1,2]. Another interest of SA relies on the capacity of getting a 3D information in a single x-y scan.

*OCIS codes:* 070.0070, 090.0090, 110.0110, 180.0180.

## 1) INTRODUCTION

Making fast 3D images with a good in-depth resolution through turbid media have always been a major issue. The problem is double with scattering media: first the scattering medium generally attenuates strongly the signal, which decreases the Signal to Noise Ratio (SNR) and second the wavefront is highly perturbed, which degrades the Point Spread function (PSF) of the imaging system and therefore the resolution. Several ways to overcome these problems have been proposed; two main methods aiming at keeping a good optical resolution are actively developed. The first one uses pre-compensation of the wavefront before propagation, to improve the resolution. This technique is used successfully both with optics or acoustic modality [3,4], but it requires *a priori* knowledge of the medium. The second one only uses ballistic photons to make images: Optical Coherence Tomography (OCT) [5], confocal [6], fluorescence [7] and nonlinear microscopy [8] belong to this family as well as tomographic diffractive microscopy [9]. Our imaging technics (LOFI), based on optical reinjection in the laser cavity also belongs to this second family [10].

In this paper we give a brief reminder of what is LOFI and how it can be used to make images with two different configurations. The first one is conventional imaging microscope that belongs to the confocal microscopes family whereas the second one is a SA based microscope. We will see that this last imaging modality has the advantage of giving access to 3D imaging in only one x-y scan (whereas conventional modality needs a very time-consuming x-y-z acquisition) and so has an important speed advantage for 3D images. In addition, we will show that SA configuration permits obtaining good resolutions with high working distance and that, in return, the drawback of SA imagery is that it has a worse photometric balance but which is partially compensated by the important sensitivity of LOFI [2]. In the first part of the paper, we remind the two imagery setups and compare their resolutions. In a second part, we provide both a theoretical and experimental comparison of photometric performances of these two configurations and we give the SNR accessible in both direct and SA deep imaging in a turbid medium. We also show that the limitation is due to parasitic reflection at the input interface of this medium. We conclude by giving two ways of improvement of both resolution and SNR of the SA configuration that will have to be explored in our future work.

## 2) Reminder on LOFI and synthetic aperture imaging

### a) Experimental setup

Figure 1 shows a description of the LOFI experimental setup [11]. The laser is a cw Nd:YVO$_4$ microchip emitting about 85 mW power at $\lambda$ = 1064 nm. This laser has a relaxation frequency near 2.5 MHz and is then frequency-shifted near this relaxation frequency. This frequency shift is chosen close to the relaxation frequency of the laser in order to increase its sensitivity to reinjected photons from imaged target. The laser beam is then sent to the bidimensional target using a galvanometric scanner constituted by two rotating mirrors, respectively called M$_x$ and M$_y$. The first one allows scanning of the target in the horizontal direction (x direction), and the second one in the vertical direction (y direction). The angular orientations of the galvanometric mirrors are given by the angles $\alpha_x$ and $\alpha_y$, respectively. For a classical - or confocal- LOFI experiment (Figure 1 a), the laser is focused in the target plane. For the SA (Synthetic Aperture) LOFI experiment (Figure 1 b), the laser is focused in front of the target plane. For the SA LOFI experiment (see Figure 1), l + d + L is the distance between the focal spot and the target plane. In both cases, the beam diffracted and/or scattered by the target is then reinjected inside the laser cavity after a second pass in the galvanometric scanner and the frequency shifter. Under the influence of reinjected photons, the laser output power is modulated at twice the frequency shift (there is two pass in frequency shifter). A small fraction of the output beam of the microchip laser is sent to a photodiode.

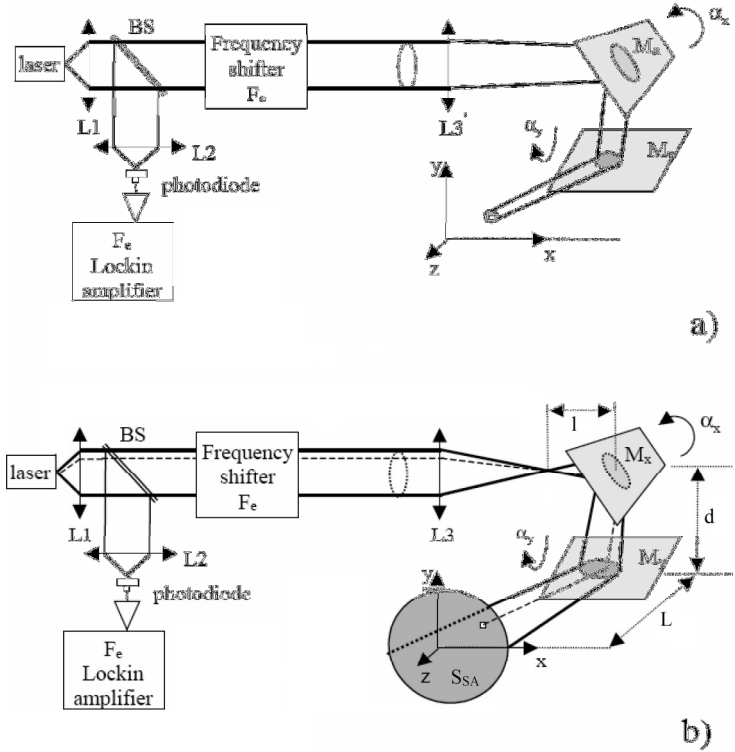

**Figure 1 :** Description of the LOFI experimental setups. The target is located in the vertical plane (x, y, z=0). L1, L2, L3 are lenses; BS is a beam splitter (T = 90%); $F_e$ is the total optical frequency shift. $M_x$ and $M_y$ are the rotating mirrors that allow scanning of the target in the horizontal direction x and the vertical direction y respectively. The angular orientations of the galvanometric mirrors are given by the angles $\alpha_x$ and $\alpha_y$. (a) Conventional LOFI experiment where the laser is focused in the target plane. (b) SA LOFI experiment. The laser is focused in front of the target plane. l is the focal spot−$M_x$ distance, d is the $M_x$−$M_y$ distance, L is the $M_y$−target plane distance, and $S_{SA}$ is the gaussian laser beam surface in the target plane.

The delivered voltage is analyzed by a lock-in amplifier, which gives the LOFI signal (i.e. the amplitude and the phase of the backscattered electric field) at the demodulation frequency $F_e$. Experimentally, the LOFI images (amplitude and phase) are obtained pixel by pixel (i.e., point by point, line after line) by full 2D galvanometric scanning ($\alpha_x$, $\alpha_y$).

### b) LOFI Signal

In the case of weak optical feedback, the coherent interaction (beating) between the lasing electric field and the frequency-shifted optical feedback field leads to an amplitude modulation of the laser output power [12,13]:

$$\Delta P_{out}(\alpha_x, \alpha_y) = 2G_R(\Omega_e)P_{out}\sum_i \sqrt{R_e(\alpha_x, \alpha_y, x_i, y_i)}\cos(\Omega_e t + \Phi(\alpha_x, \alpha_y, x_i, y_i) + \Phi_R(\Omega_e)) \quad (1)$$

where $P_{out}$ is the photon output rate (number of photon per second) and $\Omega_e = 2\pi F_e$ with $F_e$ the frequency shift. $G_R(\Omega_e)$ and $\Phi_R(\Omega_e)$ are respectively the dynamical gain and the dynamical phase shift, which only depend on the laser parameters [13]:

$$G_R(\Omega_e) = \gamma_c \frac{\sqrt{(\eta\gamma_1)^2 + \Omega_e^2}}{\sqrt{(\Omega_R^2 - \Omega_e^2)^2 + (\eta\gamma_1\Omega_e)^2}}$$
$$\Phi_R(\Omega_e) = \arctan(\frac{\Omega_e(\Omega_R^2 - \Omega_e^2 - (\eta\gamma_1)^2)}{\eta\gamma_1\Omega_R^2}) \quad (2)$$

where $\Omega_R = 2\pi F_R = (\gamma_1\gamma_c(\eta-1))^{1/2}$ is the laser relaxation frequency, $\gamma_c$ is the laser cavity damping rate, $\gamma_1$ is the population inversion decay rate, and $\eta$ is the normalized pumping parameter. Eq. (2) clearly shows a resonance when $\Omega_e = \Omega_R$, which provides the sensitivity of the LOFI technique. For our microchip laser, we have $\gamma_c \approx 7\times10^9$ s$^{-1}$, $\gamma_1 \approx 3.3\times10^4$ s$^{-1}$, and for $\eta = 1.7$ we obtain $G_R(\Omega_R) = \gamma_c/\eta\gamma_1 \approx 1.3\times10^5$. This gain allows us detecting a very weak optical feedback.

In Eq. (1), we have assumed that the target under investigation could be decomposed as a discrete sum of punctual targets indexed by i and characterized by their effective power reflectivity $R_e(\alpha_x,\alpha_y,x_i,y_i)$ depending on the laser incidence and by the optical phase shift $\Phi(\alpha_x, \alpha_y, x_i, y_i)$ due to the optical round trip between the laser and the punctual target. Eq. (1) also shows that the optical feedback is formed by the coherent interaction (i.e. addition) of each punctual target point illuminated by the gaussian laser beam spot. The demodulation of the laser power at the frequency shift $F_e$ by the means of a lock-in amplifier gives us the quadrature components of the LOFI signal:

$$I(\alpha_x, \alpha_y) = 2G_R(\Omega_e)P_{out} \sum_i \sqrt{R_e(\alpha_x, \alpha_y, x_i, y_i)} \cos(\Phi(\alpha_x, \alpha_y, x_i, y_i))$$
$$Q(\alpha_x, \alpha_y) = 2G_R(\Omega_e)P_{out} \sum_i \sqrt{R_e(\alpha_x, \alpha_y, x_i, y_i)} \sin(\Phi(\alpha_x, \alpha_y, x_i, y_i))$$
(3)

and thus the complex expression of the LOFI signal:

$$h(\alpha_x, \alpha_y) = I(\alpha_x, \alpha_y) + jQ(\alpha_x, \alpha_y)$$
$$= |h(\alpha_x, \alpha_y)| \exp(j\Phi_S(\alpha_x, \alpha_y))$$
(4)

We must now consider two possibilities:

- i = 1 (one pixel corresponds to one image point): it corresponds to conventional LOFI where we scan the object with a focused beam. We can get an amplitude [14,15,16] |h($\alpha_x$,$\alpha_y$)| or phase [17,18,19,20] image $\Phi_S(\alpha_x,\alpha_y)$.

- i >> 1 (one pixel contains informations of the whole field): it corresponds to an image acquired by a defocused beam. This raw complex image h($\alpha_x$,$\alpha_y$) must be filtered to realise a post numerical focusing. This imaging technique is called Synthetic Aperture (SA) LOFI [11].

In the following, we will index all parameters related to the conventional setup with "c" and those related to SA imaging with "SA".

## c) Point Spread Function (PSF)

### i) Conventional imaging

With the conventional configuration, the resolution is simply given by the Gaussian beam waist. Considering that the LOFI signal gives access to the electric field amplitude, we simply have a PSF signal (image of a punctual target) given by:

$$|h_c(x, y)| \propto \exp\left[-2\left(\frac{x^2 + y^2}{RES_c^2}\right)\right] \qquad (5)$$

The resolution is:

$$RES_c = \sqrt{2}\, r_c = \frac{\sqrt{2}\lambda}{\pi W_c} f_c \qquad (6)$$

where $\lambda$ is the wavelength of the laser, $r_c$ is the beam waists in the target plane, $W_c$ is the gaussian radius of the beam (in the x and y directions) before the lens L3' (see Figure 2) and $f_c$ is the focal length of L3'.

### ii) Synthetic imaging

Synthetic Aperture consists in scanning the target with a diverging beam while recording amplitude and phase informations on the movement of the laser spot with respect to the target. It enables realising a numerical focalization and recovering a good resolution.

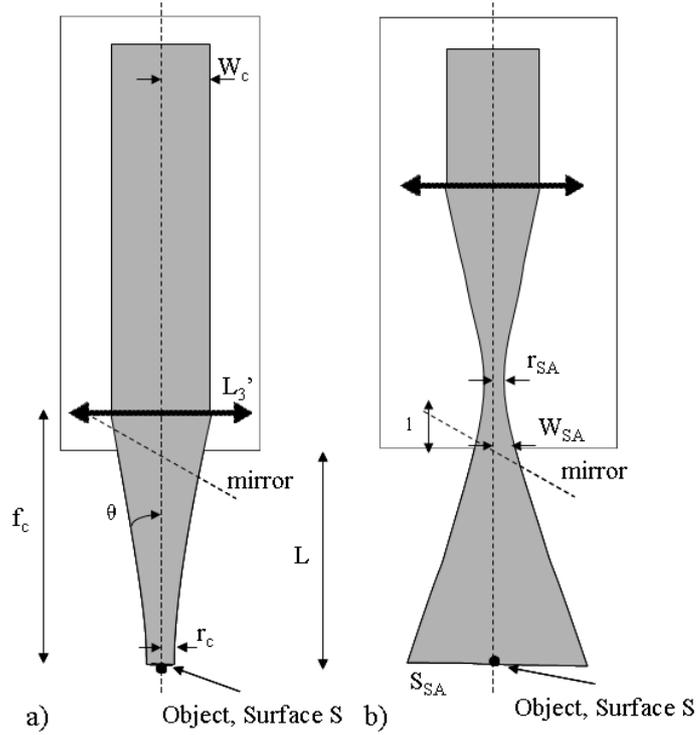

**Figure 2 : Sketch of the two configurations discussed in the paper. a) Conventional LOFI : $W_c$ is the beam waist before L3', $r_c$ is the waist in the target plane and $\theta$ the numerical aperture. b) SA LOFI : $r_{SA}$ is the beam waist before the galvanometric mirrors, l is the distance between them, $S_{SA}$ is the surface of the beam in the object plane. L is the distance between the last optical element (the scanning mirror) of the setup and the target of surface S.**

The technique was introduced first in SAR imaging (Synthetic Aperture Radar) [21,22] to overcome the fact that no large portable aperture component exists for radio waves. Then it has been applied to optical wavelengths with $CO_2$ [23] and Nd:YAG microchip laser source [24,25] in what has been called SAL (Synthetic Aperture Laser). Here we propose a scanning of the target with galvanometric mirrors; it has the advantage of being vibration noise free and easy to implement. In this condition, it has already been shown [11] that the PSF is given by:

$$|h'_{SA}(x,y)| \propto \exp\left[-2\left(\frac{x^2}{RES_{SA,x}^2} + \frac{y^2}{RES_{SA,y}^2}\right)\right] \quad (7)$$

where the resolution is given by (see Figure 1 for the parameters):

$$RES_{SA,x} = \frac{r_{SA}(d+L)}{l} \text{ and } RES_{SA,y} = \frac{r_{SA}L}{l+d} \quad (8)$$

In this expression $r_{SA}$ is the beam waist of the laser in the SA (see Figure 2).

### iii) Discussion

It is now possible to compare transverse resolution of conventional and SA LOFI. We can see from Eq. (6) and Eq. (8) that these two configurations are equivalent with regard to the resolution. To compare them, we simplify the expressions by assuming that d ≈ 0 and we define the working distance L as the distance between the last optical element ($M_y$ mirror) and the target. In the conventional LOFI imaging since the galvanometrics mirrors are very closed to the focusing lens (see Figure 2 a), we assume L ≈ $f_c$. As a result, the expressions of resolution are simplified:

$$RES_c = \frac{\sqrt{2}\lambda}{\pi W_c} f_c \approx \frac{\sqrt{2}\lambda}{\pi W_c} L$$

$$RES_{SA} = \frac{r_{SA}}{l} L = \frac{\sqrt{2}\lambda}{\pi (\frac{\sqrt{2}\lambda l_x}{\pi r_{SA}})} L = \frac{\sqrt{2}\lambda}{\pi W_{SA}} L \quad (9)$$

We can observe that the resolution degrades with observation distance L in both cases and that in terms of lateral resolution, the synthetic setup is strictly equivalent to a conventional setup with an equivalent lens diameter equal to L3': $W_{SA}$ = (√2 λ l) / (π $r_{SA}$), which corresponds to the gaussian radius of the laser beam on the galvanometric mirrors, multiplied by a factor √2 (see Figure 2).

In both cases (conventional and synthetic microscope), we have a limitation: in the conventional imaging, it comes from the diameter of L3' (the galvanometric mirrors are usually larger than the diameter of microscope objective L3') whereas in the synthetic setup, it corresponds to the size of the galvanometric mirrors.

The advantage of the synthetic imaging is that we can make aberration free images with both a large numerical aperture (*i.e.* a good resolution) and with an important working distance

($W_{SA}$ is usually larger than microscope objectives radius). This leads to the possibility of obtaining resolved images with an important working distance. Another important thing is that the use of microscope objective which are expensive is not necessary.

## 3) Experimental example

We give an example of image that can be made by using LOFI principle. The target observed is a piece of PVC with a 1 mm diameter aperture in front of a reflective layer made of silica beads (40 μm diameter). The classical bright field transmission microscope image is presented in Figure 3.

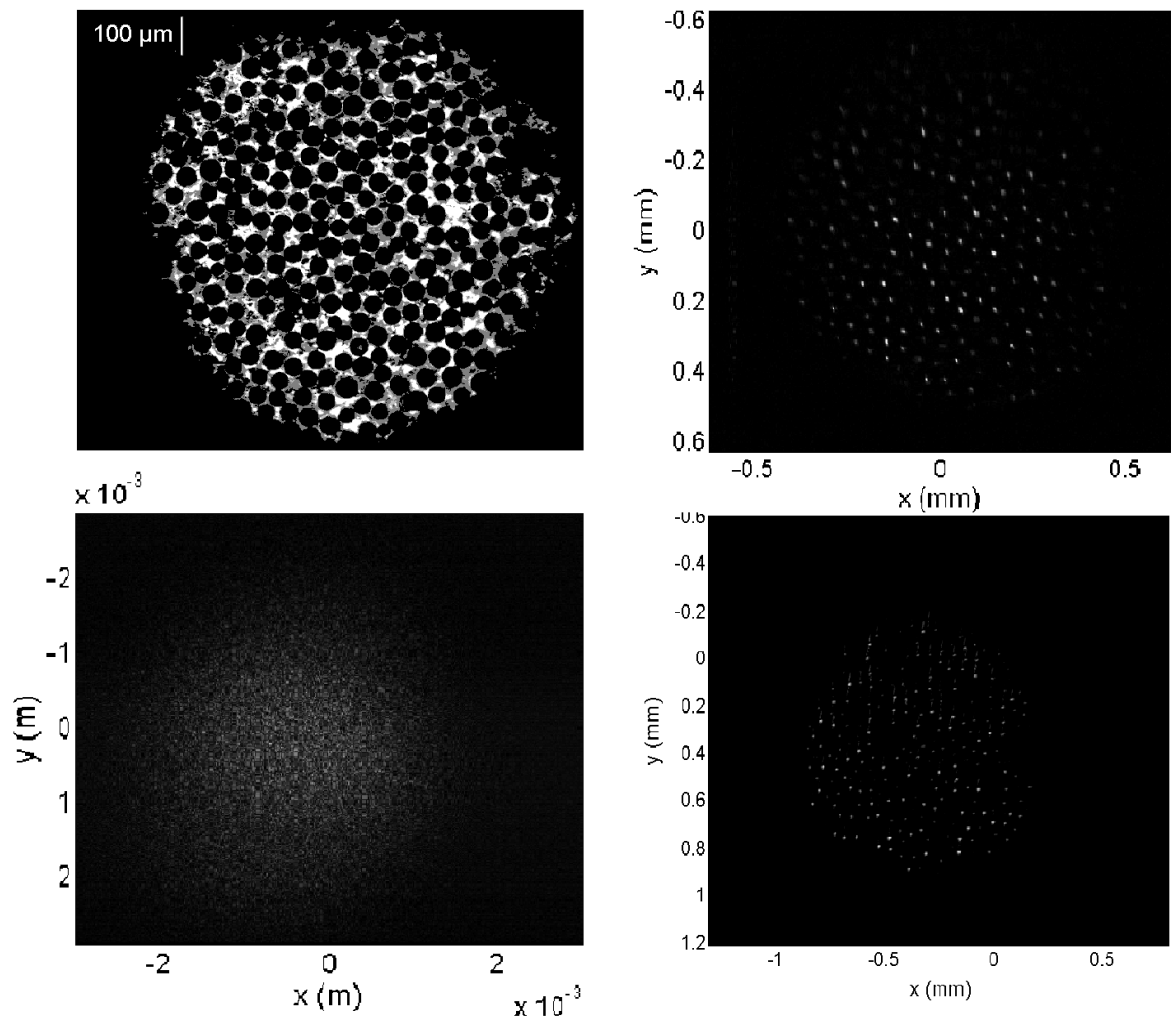

**Figure 3: a) Object under microscope. It is composed of reflective silica beads of 40 μm diameter behind a circular aperture of 1 mm diameter. The bright field transmission image is made through a Zeiss**

**microscope objective with a magnification of 10 and a 0.25 numerical aperture (focal length of 20 mm). b) LOFI amplitude image of the target (size: 512*512 pixels). The image is formed through the same Zeiss objective but with a laser beam input size of $W_c$ = 1.3 mm, a resolution $RES_c$ of ~7 μm is expected which is coherent with bead's image size on the image. A laser power of 2 mW is sent on the target. c) Raw image of the target (size: 2048*2048 pixels). Parameters are: $r_{SA}$ = 20 μm, l = 10 cm, d = 1 cm, L = 2.5 cm. The power sent on the target is 1.5 mW. The beam size on the target plane is equal to 1.7 mm: beads are not resolved and the size of raw image is enlarged comparing to real image. d) Synthetic image after filtering of the raw image. Predicted resolution is $RES_{SA,x}$ = 5.7 μm in X direction and $RES_{SA,y}$ = 5.3 μm in Y direction which is coherent with bead's image size on the image. We found the good focusing plane (so in the target plane) by using the detection criteria described by F. Dubois [26], that is this algorithm that will be used each time synthetic aperture filtering is performed. The image in the lower right corner has the same size the image in the lower left corner but is zoomed on the object to be comparable to conventional image of the image in the upper right corner.**

This object is then imaged by a conventional LOFI microscope; the image is shown in Figure 3.

We can observe that images of the beads are smaller than their real size, which can be explained by as the beads are spherical, all reinjected photons seems to be reflected from the center of the beads and they behave like a Dirac-like reflector. So, the size of the beads images corresponds to the PSF. Finally, Figure 3 gives the image of the objet by the synthetic microscope before and after adapted filtering.

In both cases, the resolution gives us the possibility to resolve the silica beads. The differences between the sizes of images (1.2*1.2 mm for the conventional image and 6*6 mm for the SA image) can be explained easily. As the raw image of Figure 3 is in fact a larger defocused image of the object, one needs to scan with a larger angle (a factor ~4 here), which explains the larger size of the raw figure (comparing to the conventional case) required. Image of Figure 3 of the object after filtering has the same size that before filtering but it simply zoomed to be easily compared to conventional image.

Compared to conventional imaging, another interest of the synthetic configuration is the possibility to get a 3D information with a single x-y scan followed by a numerical focalization in the different z planes whereas with a conventional setup, it is necessary to make a x-y scan in each focalization plane. So there is an important gain in acquisition time for 3D imaging. Figure 4 gives an example of images of a pseudo 3D object in different planes, this object is composed of one 350 μm width horizontal strip which is 4 cm after $M_y$ and of 3 double slits spaced by respectively 400, 600 and 800 μm at a distance of 9 cm after $M_y$.

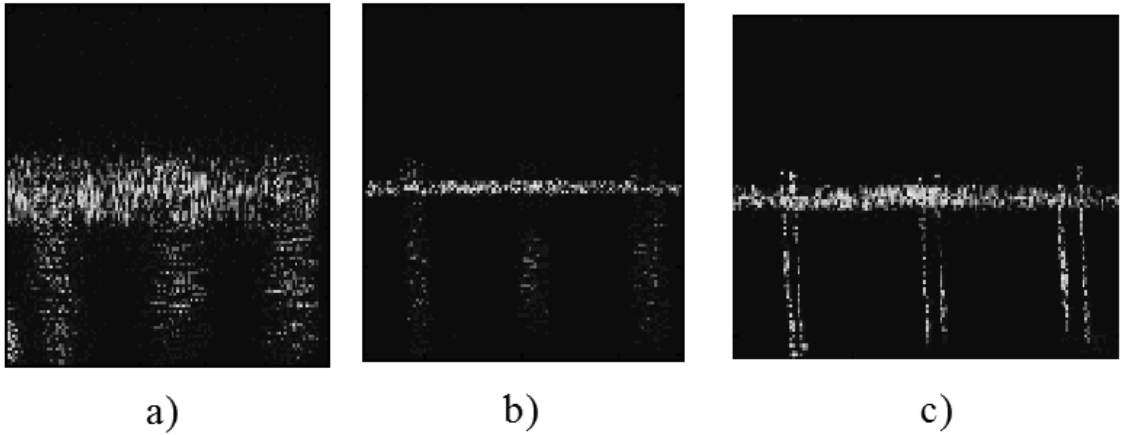

**Figure 4: Image of a 3D object composed of one 350 μm width horizontal strip which is 4 cm after $M_y$ and of 3 double slits spaced by respectively 400, 600 and 800 μm at a distance of 9 cm after $M_y$. a) gives raw image, the beam size in the strip plane is 1.4 mm and 2.2 mm in the slits plane, slits are not resolved. b) the image numerically focused on the strip plane (double slits are not resolved), theoretical resolution is 20 μm. c) is the image numerically focused on double slits which are now resolved (theoretical resolution is 40 μm).**

## 4) Photometric budget of conventional and synthetic LOFI

Our goal here is to evaluate and compare the sensitivity of the two imaging systems. More particularly, the influence of several parameters on the backscattered flux by the target is calculated and validated by experimental measurements and the two configurations are compared. In order to separate difficulties and to simplify the problem, we choose experimental parameters where the resolution is comparable between the two setups.

Taking $l = 5$ cm, $d = 1$ cm, $W_c \approx 0.9$ mm, $W_{SA} \approx 1.3$ μm, we get $RES_c / RES_{SA} \approx 1.45$. In this case, the resolution is similar for both configurations. We take, in addition $d \approx 0$ as $d \ll l$ and L and so expressions are simplified without big mistakes.

## a) Theoretical sensitivity

We consider the situation of a lambertian diffusive target with a surface S. From Eq. (1) we see that the LOFI signal from the object i is proportional to $\sqrt{R_e(x_i,y_i)}$ with $R_e$ the intensity reflection coefficient. So the signal is proportional to the amplitude of the reinjected field in the cavity. In the photometric analysis, what is important is the number of photons present in an image. This is why, in what follows, we work on the signal power in pixels that is to say, the square of the LOFI signal.

The goal is now to calculate the power backscattered by a diffusive object of albedo $\rho$ and surface S smaller than the PSF surface (non-resolved object).

### i) Conventional LOFI

This is the easiest configuration to calculate, the signal power being simply concentrated in one pixel and no mathematical treatment is applied. Figure 2 a shows the scheme of conventional LOFI. The mean illuminance in the focal plane is given by:

$$E_c = \frac{P_{out}}{\pi r_c^2} = \frac{P_{out} \pi W_c^2}{\lambda^2 L^2} \tag{10}$$

where $P_{out}$ is the laser input power and $\lambda$ the wavelength of the laser. The geometric extent $G_c$ and the luminance $L_c$ in the object plane are given by:

$$G_c = S\pi\theta^2 = \frac{S\pi W_c^2}{L^2} \tag{11}$$

$$L_c = \frac{\rho E_c}{\pi} \tag{12}$$

In Eq. (12), $E_c$ represents the illuminance in the object plane. From Eq. (10), Eq. (11) and Eq. (12), we can finally deduce the reinjected input power:

$$P_c = G_c L_c = \frac{S\pi W_c^2}{L^2} \rho \frac{P_{out} \pi W_c^2}{\pi \lambda^2 L^2} = \rho P_{out} S \frac{\pi W_c^4}{\lambda^2 L^4} \quad (13)$$

From this formula, we expect the signal power to show a quick decay with the observation distance L and on the contrary an improvement with $W_c$, that is to say the aperture of the focusing lens L3'.

### ii) SA LOFI

In this configuration (Figure 2 b), we are in a more complex situation: a pixel in the synthetic image is reconstituted from several pixels of the raw image. Consequently, the total power in the PSF of the synthetic image is:

$$P_{SA} = \frac{N_{pixels} P_{pixelSA}}{\sqrt{2}} \quad (14)$$

where $N_{pixels}$ is the number of pixels contained in the PSF of the raw image and $P_{pixelSA}$ is the mean power in each raw pixel. The factor two at the denominator is due to the fact that the adapted filter maximises the SNR but eliminates half of the raw signal power. $N_{pixels}$ and $P_{pixelSA}$ are given by (see Figure 2 for parameters definition):

$$N_{pixels} = \frac{S_{SA}}{S_{pixel}} = \frac{\pi(\frac{\lambda}{\pi r_{SA}}(l+L))^2}{\pi(\frac{r_{SA}L}{l})^2} = \frac{\lambda^2(l+L)^2 l^2}{\pi^2 r_{SA}^4 L^2} \quad (15)$$

Here, $S_{SA}$ and $S_{pixel}$ represent respectively the surface of the defocused beam in the object plane and the surface of the pixel, which is chosen to correspond to the surface of the PSF.

$$P_{pixelSA} = G_{SA} L_{SA} \tag{16}$$

with $L_{SA}$ the luminance in the object plane and $G_{SA}$ the geometric extent in this configuration. We have analogous result as direct focalisation LOFI, Eq. (11) and Eq. (12) simply become:

$$G_{SA} = \frac{\pi r_{SA}^2 S}{(l+L)^2} \tag{17}$$

$$L_{SA} = \rho \frac{E_{SA}}{\pi} = \frac{\rho}{\pi} \frac{P_{out}}{\pi(\frac{\lambda}{\pi r_{SA}}(l+L))^2} = \rho P_{out} \frac{r_{SA}^2}{\lambda^2 (l+L)^2} \tag{18}$$

where $E_{SA}$ is the illuminance in the target plane. Finally, from Eq. (15), Eq. (16), Eq. (17) and Eq. (18), we get the power in the PSF synthetic pixel:

$$P_{SA} = \frac{1}{\sqrt{2}} \frac{\lambda^2 (l+L)^2 l^2}{\pi^2 r_{SA}^4 L^2} \frac{\pi r_{SA}^2 S}{(l+L)^2} \rho P_{out} \frac{r_{SA}^2}{\lambda^2 (l+L)^2} = \rho P_{out} \frac{l^2 S}{\sqrt{2} \pi L^2 (l+L)^2} \tag{19}$$

### iii) Comparison

From the Eq. (19), two possibilities must be distinguished.

The first is when $l \ll L$, which corresponds to a target far from the imaging setup (telemetry for instance). In this case, Eq. (19) becomes:

$$P_{SA, l \ll L} = \rho P_{out} \frac{l^2 S}{\sqrt{2\pi} L^4} \tag{20}$$

As for the resolution, the two photometric balances (Eq. 13 and Eq. 20) show that conventional and SA setups have the same dependence with the observation distance L, that is in $1/L^4$ for the photometry. The other important aspect is the comparison of the

proportionality constant in front of this dependence. The ratio of these coefficients is given by:

$$\frac{P_c}{P_{SA,l<<L}} = \frac{\sqrt{2}\pi^2 W_c^4}{\lambda^2 l^2} \quad (21)$$

Numerically, if we take the parameters indicated at the beginning of this section, which correspond to the same resolution, we get from Eq. (21) a photometric ratio of about 2800 in favour of direct focusing. This big difference in sensitivity can easily be explained by reminding that LOFI is in fact a confocal microscope (the coupling of the reinjected photons with the stationary laser transverse mode plays the role of the pinhole) whereas in SA imaging, we collect photons that were scattered far from the equivalent pinhole-conjugated zone (which is in the $r_{SA}$ plane). With the conventional LOFI configuration, this plane coincides with the target and therefore we get a maximum collection of backscattered photons.

Concerning the acquisition time required to make the image of one Dirac-like object, it only corresponds to one pixel integration time T (here 100 μs) in the case of conventional imaging, whereas for the synthetic image, this time is multiplied by the number of pixels $N_{pixels,l<<L}$ used in the filtering (Eq. (15)):

$$T_{acq,SA,l<<L} = N_{pixels,l<<L} T \approx \frac{\lambda^2 l^2}{\pi^2 r_{SA}^4} T \quad (22)$$

For instance, with parameters used here, we get $T_{acq,SA}$ = 180 ms which is not realistic because of the speed of the galvanometric mirrors (we should add the displacement time of galvanometric mirrors). This time increases if we want to change parameters to ameliorate the resolution ($r_{SA}$ decrease or l increase). There is an exception for the parameter L, which does not influence the measurement time.

The second possibility is when l >> L (i.e. target very close to the imaging dispositive). It corresponds to microscopy, which is our main interest. In this case, Eq. (19) becomes:

$$P_{SA,l>>L} = \rho P_{out} \frac{S}{\sqrt{2\pi}L^2} \qquad (23)$$

Here we get:

$$\frac{P_c}{P_{SA,l>>L}} = \frac{\sqrt{2}\pi^2 W_c^4}{\lambda^2 L^2} \qquad (24)$$

The closer is the object, the less efficient is synthetic imaging compared to the conventional mode. The photometric efficiency of the synthetic microscope does not depend on parameters $r_{SA}$ and l.

Concerning the acquisition time $T_{acq,SA,L<<l}$ to measure the raw image of a Dirac-like target, it is given by:

$$T_{acq,SA,l>>L} = N_{pixels,l>>L}T \approx \frac{\lambda^2 l^4}{\pi^2 r_{SA}^4 L^2}T \qquad (25)$$

For instance, with parameters used here, and L = 1 cm for example, we get $T_{acq,SA}$ = 4.5 s which is much bigger than in case L >> l. This measurement time increases when the PSF size decreases.

In both cases (L >> l and L << l), we see that measurement time increases with the improvement of the resolution contrary to conventional imaging. However an important remark is that if we want to image an object distributed on N close pixels (the calculation concerns the image of one Dirac), the added necessary time is only (N-1)T as in the conventional case and not $(N-1)T_{acq,SA}$. Bigger is the image and smaller is the difference between acquisition times of both setups.

## b) *Experimental measurements*

To confirm these theoretical results, we have realised the two microscopes and measured the signal power at different working distances; the target is the object of Figure 3. As we said before, the power in one pixel is defined as the square of the voltage from the lock-in amplifier; units are arbitrary but the point is to keep the same unit for all the measurements.

There are three things to verify for validating our models: first, the $1/L^4$ law of the conventional setup, second the $1/L^2(l+L)^2$ law of the synthetic microscope and third the photometric ratio of 2800 when $L \gg l$. For that purpose, we make images of the object described on Figure 3 at different distances L with the two setups. Curves of Figure 5 correspond to the power in the image versus L (power is calculated by simply summing the square of the voltage recorded in each pixel). A theoretical fit is made with $1/L^2(l+L)^2$ for the synthetic microscope and with $1/L^4$ for the conventional setup. We take parameters corresponding to comparable resolutions (see the introduction of this section).

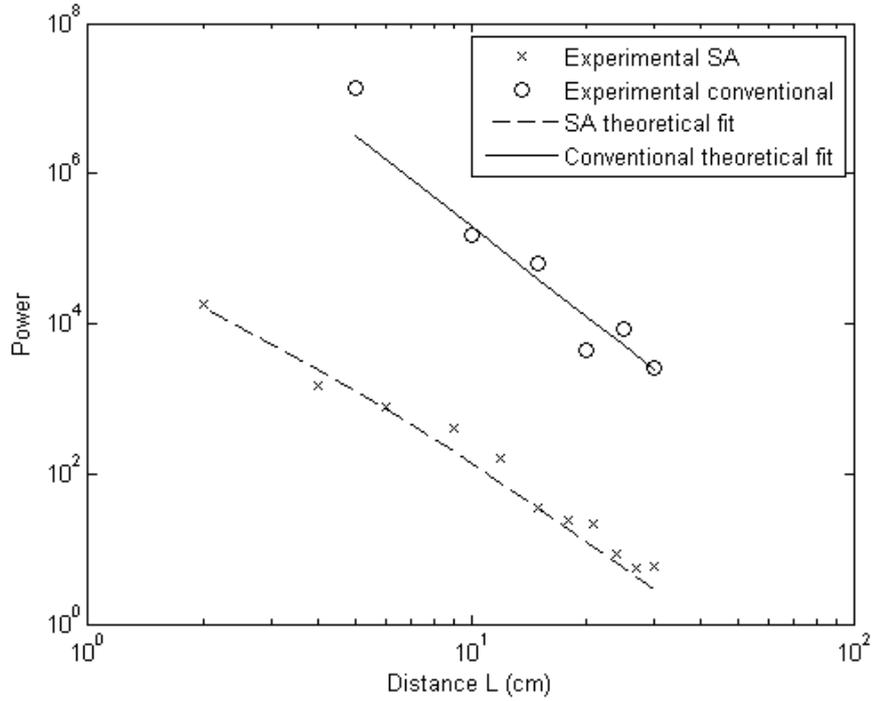

**Figure 5: Experimental photometric comparison between conventional and SA LOFI with same resolutions. Parameters are: $W_c$ = 0.9 mm, $W_{SA}$ = 1.3 mm, $r_{SA}$ = 20 μm, l = 5 cm, d = 1 cm. The laser power after frequency shifter is 2 mW (the power sent on the beads is reduced comparing to the 50 mW avalaible power to stay in weak reinjection conditions). For the conventional microscope, the focal length of L3' (L) is varied from 5 to 30 cm. The conventional photometric balance is fitted with $1 / (l + L)^2 L^2$ whereas the SA photometric balance is fitted to $1/L^4$.**

These curves show that the theoretical dependence of the signal power with L is experimentally confirmed. Moreover, there is a factor 1000 between the fit curves for L >> l, which approximately matches with the expected factor of 2800 (the calculation is an approximation: we have neglected the astigmatism, and the non-lambertian character of the object).

## 5) Accessible performances and limitations

In this section, we come back to our initial goal: obtain in-depth image through turbid medium. We consider an experimental setup where the target is in a tank filled with water (Figure 6). Milk is used at various concentrations as a scattering medium. We measure the maximum milk concentration one can reach, before the target becomes not distinguishable from the background (SNR = 1). Then we are able to completely compare (taking into account the background) the performances of the two configurations by getting the ratio of the two SNR. An experimental comparison with the shot noise limitation is also given.

### a) Theoretical predictions

In an experimental configuration such as described in Figure 6, the background noise and the resolution perturbations necessarily comes from parasitic reflections on elements after the mirrors since the flux of photons reflected before the mirrors is constant and can be easily numerically filtered. More specifically, there are three possible parasitic sources: two diffusive elements: the input face of the tank and the scattering fatty particles in the milk and one refractive element: the water of the milk of index n = 1.33. As we said, these different elements can have effects on both resolution and SNR of final image of the object. More precisely, the two diffusive elements have only an effect on the SNR since only photons which are diffused by the object participate in the image formation while photons diffused by milk or entry face of the tank haven't a coherent phase and so create only additive noise; the resolution is not affected. For the water of the milk it is the contrary, the SNR is not affected whereas the resolution is slightly changed. Indeed, using Fermat theorem, it can be shown that when we add a homogeneous medium of index n in front of the object, we must consider its image through the medium. Thus we must use $L_{eq} = L(1 - d/n)$ instead of L where d is the distance between the entry face of the tank and the object and so the resolution is improved by a factor $L_{eq}/L$ exactly like in immersion objectives. This effect is negligible in what follow and we will consider only the effect on SNR.

To compare the performances of the conventional and synthetic configurations, it is important to compare their respective SNR ratios instead of the signals powers, which gives only one part of the information. In order to make a theoretical calculation of this SNR ratio, we consider a parasitic reflection on the tank with an albedo $\rho_{tank}$. By doing so, parasitic reflections on mirrors are neglected and milk scattering is included in an effective albedo $\rho_{tank}$. This last approximation is justified by the fact that the laser beam is quickly attenuated in the milk and so that the parasitic reflections mainly come from the input face of the tank.

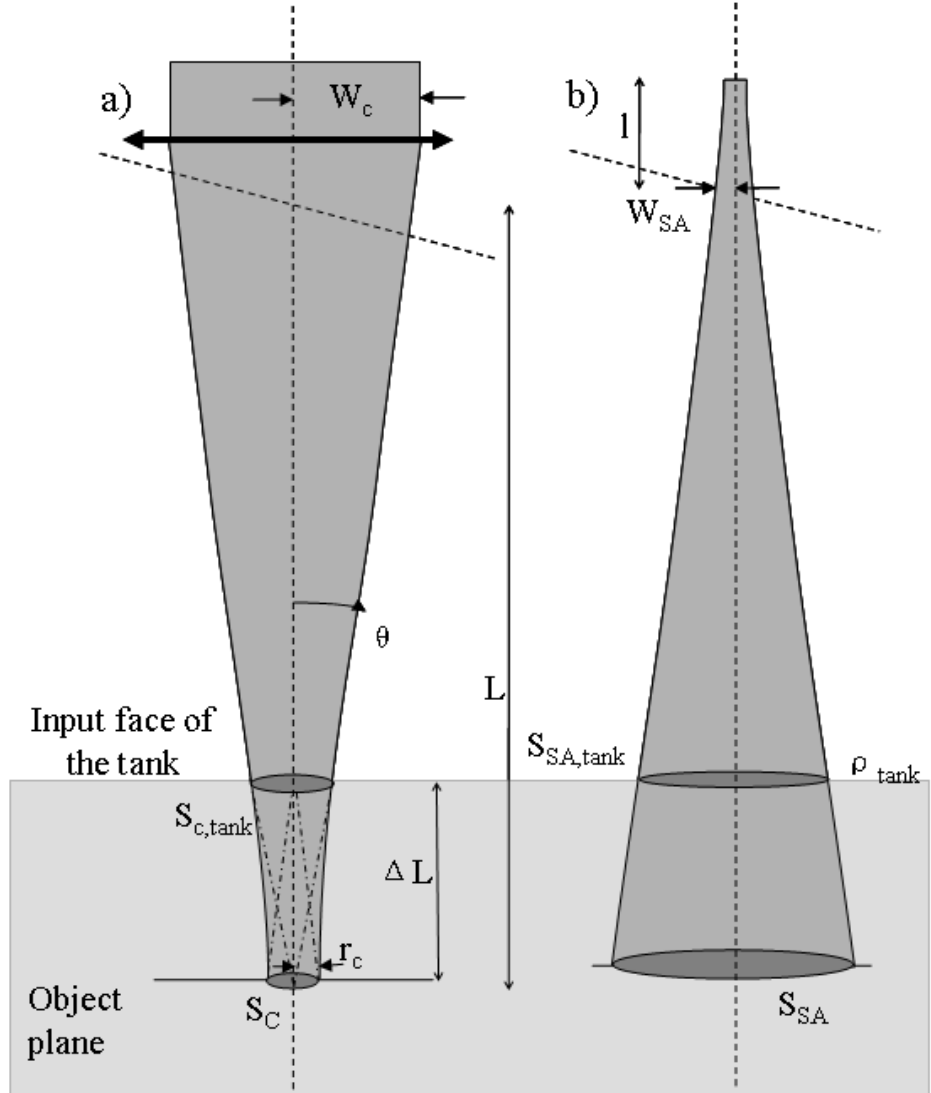

**Figure 6: Scheme of the experimental setup. a) conventional configuration, b) SA LOFI configuration. The object of Figure 3 is placed in a tank filled with milk diluted in water. $\Delta L = 4$ cm is the distance between the input face of the tank and the object (*i.e.* the distance travelled by the laser beam in the solution). Experimental parameters are: $W_c = 0.9$ mm, $W_{SA} = 1.3$ mm, $S_c = 5.6.10^{-9}$ m$^2$, $S_{c,tank} = 3.2.10^{-7}$ m$^2$, $S_{SA} = 3.1.10^{-5}$ m$^2$, $S_{SA,tank} = 1.88.10^{-5}$ m$^2$, $r_{SA} = 20$ µm, $l = 5$ cm, $f'(L3) = L = 12$ cm, power sent on the target is 2 mW. Milk concentrations are chosen from 0 to 2.5 % in volume.**

In Eq. (13) and Eq. (14) the signal power $P_c$ and $P_{SA}$ have been calculated from the target, which has a finite surface S and an albedo $\rho$. Our goal is to determine the background signal power $P_{c,tank}$ and $P_{SA,tank}$ from the input face of the tank, which has an albedo $\rho_{tank}$ and an

"infinite" surface. To make the calculations, we consider the fictive situation where the input face of the tank is in the target plane, the background powers are then $P_{c,tank}'$ and $P_{SA,tank}'$. From Eq. (13) and Eq. (14) and taking $S = S_c$ (surface of the tank in a pixel) and $\rho = \rho_{tank}$, assuming that we have $L \gg l$, we get:

$$\frac{P_{c,tank}'}{P_{SA,tank}'} = \frac{\rho_{tank} P_{out} S_c \frac{\pi W_c^4}{\lambda^2 L^4}}{\rho_{tank} S_c P_{out} \frac{l^2}{\sqrt{2\pi}L^4}} = \frac{\sqrt{2}\pi^2 W_c^4}{\lambda^2 l^2} \qquad (26)$$

In this expression, $S_c = \pi r_c^2$ is the surface of the beam in the conventional setup in the target plane. This ratio is equal to ~2800 with the parameters used previously (corresponding to comparable resolutions). So we now simply have to determine relations between $P_{c,tank}$ and $P_{c,tank}'$ and between $P_{c,tank}'$ and $P_{SA,tank}'$. We give the calculation of the first ratio ($P_{c,tank} / P_{c,tank}'$), the other one being very similar.

The power from the entry plane is given by:

$$P_{c,tank} = L_{c,tank} G_c = \frac{\rho_{tank} E_{c,tank}}{\pi} G_c = \frac{\rho_{tank} P_{out}}{\pi S_{c,tank}} \frac{S_{c,tank} S_c}{\Delta L^2} = \frac{\rho_{tank} P_{out} S_c}{\pi \Delta L^2} \qquad (27)$$

In this equation, $L_{c,tank}$ and $E_{c,tank}$ are the luminance and the illuminance of the laser in the input plane of the tank respectively. The signal power from the fictive situation is:

$$P_{c,tank}' = L_{c,tank}' G_c = \frac{\rho_{tank} E_{c,tank}'}{\pi} G_c = \frac{\rho_{tank} P_{out}}{\pi S_c} S_c \pi \theta^2 = \rho_{tank} P_{out} \theta^2 \qquad (28)$$

In this equation, $L_{c,tank}'$ and $E_{c,tank}'$ are the luminance and the illuminance of the laser in the plane of the target in the fictive situation.

Parameters $\Delta L$ and $\theta$ can be rewritten in the form:

$$\pi(\frac{\lambda \Delta L}{\pi r_c})^2 = S_{c,tank} \Rightarrow \Delta L = \sqrt{\frac{S_{c,tank} \pi r_c^2}{\lambda^2}} = \frac{\sqrt{S_{c,tank} S_c}}{\lambda}$$

$$\theta = \frac{\lambda}{\pi r_c} = \frac{\lambda}{\sqrt{\pi S_c}}$$

(29)

By combining Eq. (27), Eq. (28) and Eq. (29), we finally get:

$$\frac{P_{c,tank}'}{P_{c,tank}} = \frac{S_{c,tank}}{S_c}$$

(30)

By making similar calculations, it is easy to show the analogous relation:

$$\frac{P_{SA,tank}'}{P_{SA,tank}} = \frac{S_{SA,tank}}{S_{SA}}$$

(31)

Finally, the two SNR and their ratio simply expresses as:

$$SNR_c = \frac{P_c}{P_{c,tank}} = \frac{P_c}{P_{c,tank}'} \cdot \frac{S_{c,tank}}{S_c}$$

$$SNR_{SA} = \frac{P_{SA}}{P_{SA,tank}} = \frac{P_{SA}}{P_{SA,tank}'} \cdot \frac{S_{SA,tank}}{S_{SA}}$$

$$\frac{SNR_c}{SNR_{SA}} = \frac{S_{c,tank} S_{SA}}{S_c S_{SA,tank}}$$

(32)

Using parameters of Figure 6, $S_c = 5.6.10^{-9}$ m$^2$, $S_{c,tank} = 3.2.10^{-7}$ m$^2$, $S_{SA} = 3.1.10^{-5}$ m$^2$, $S_{SA,tank} = 1.88.10^{-5}$ m$^2$, the SNR ratio is expected to be equal to 68. We can thus observe that the difference between the two configurations is less important than the simpler previous power analysis showed (power ratio of 2800). Therefore taking care of noise is very important regarding the comparison between conventional and Synthetic Aperture LOFI.

### *b) Experimental results*

### i) Limitation by parasitic echoes

To validate the previous photometric analysis, we realise the two microscopes showed in Figure 6 with the parameters indicated in the legend. The target is still the object in Figure 3 immerged in water diluted milk at 4 centimetres behind entry face of the tank. The milk concentration is increased progressively from 0 to 2.5 % in volume (from 0 to 15 ml of milk in 600 ml of water) and an image is taken every 1 ml of milk added. The powers of each target image and of the background are measured and results are shown in Figure 7.

We find that there is a factor 3000 between powers of synthetic and conventional images in conformity to the theory (~2800).

In the second plot at optical density (OD) 0.8 (1.2 % vol. of milk trough 4 cm), the theoretical factor 68 between the two SNR is confirmed. This validates and confirms that the limiting background is due to parasitic reflections on the input face of the tank and on the milk. SNR ratios are not valid for all milk concentrations: at low concentration, we get signal from the support of the target so the SNR saturates. At high concentration, the signal is below the background and so the SNR is below one. Because of these limitations, the calculation of the ratio between the two SNR is valid only at concentrations from OD 0.5 to 1.

To conclude on the performances, we reach a SNR of one for a milk concentration corresponding to an OD of 1.75 (3.5 if we consider round trip) with conventional imaging and only an OD of 1.4 (2.8 in round trip) for synthetic aperture imaging.

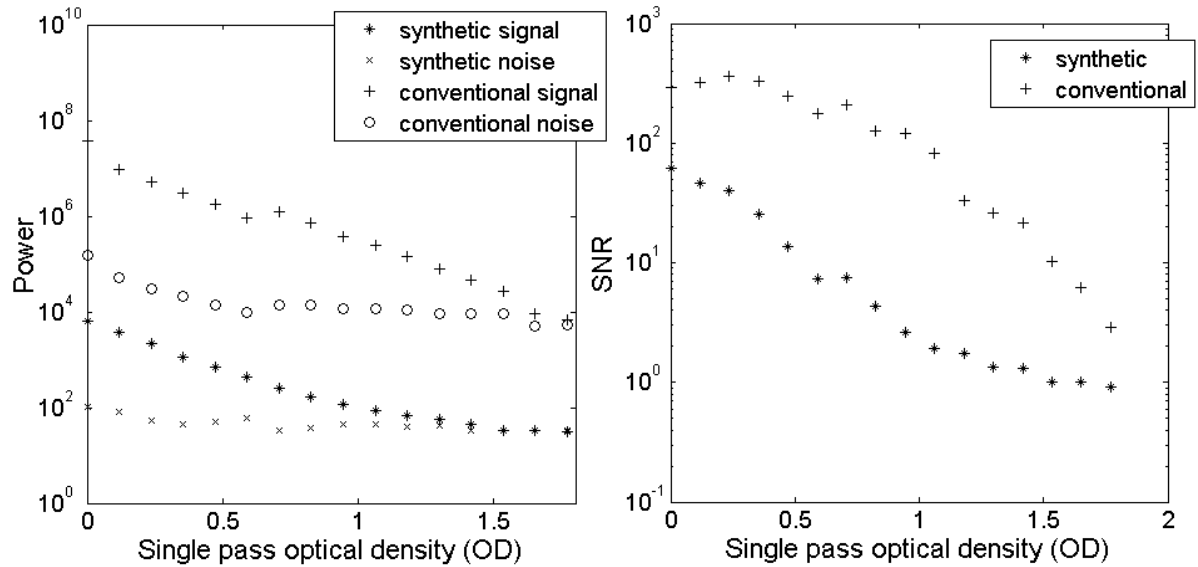

**Figure 7:** a) Experimental results associated to setups of Figure 6. Top: power reflected by the target (signal) and by the background (noise) in conventional and SA configurations versus the optical density (OD) of a diluted milk solution through 4 cm; concentrations evolve from 0 to 2.5% in volume. Bottom: plot of the SNR accessible in conventional and SA configurations versus the OD.

We can see that there is a good accordance between the energy loss and the milk attenuation: when the single pass optical density is increased by one, the power is divided by 100.

### ii) Comparison to the shot noise

It is interesting to compare the laser quantum noise [27,28] to our actual limitation by parasitic reflection on milk and input face, since shot noise is the ultimate limitation we could reach. To compare our signal to quantum noise and not be anymore limited by parasitic reflections after galvanometric mirrors, we simply realise the attenuation not by using the turbid medium but by adding an optical attenuator before the acousto-optic deflectors. As a result, the parasitic reflections after the mirrors are attenuated too and the quantum noise is finally our limitation. The target is the same as in Figure 3 and is immerged in pure water. The parameters of the microscopes are unchanged; the experimental measurements are presented on Figure 8.

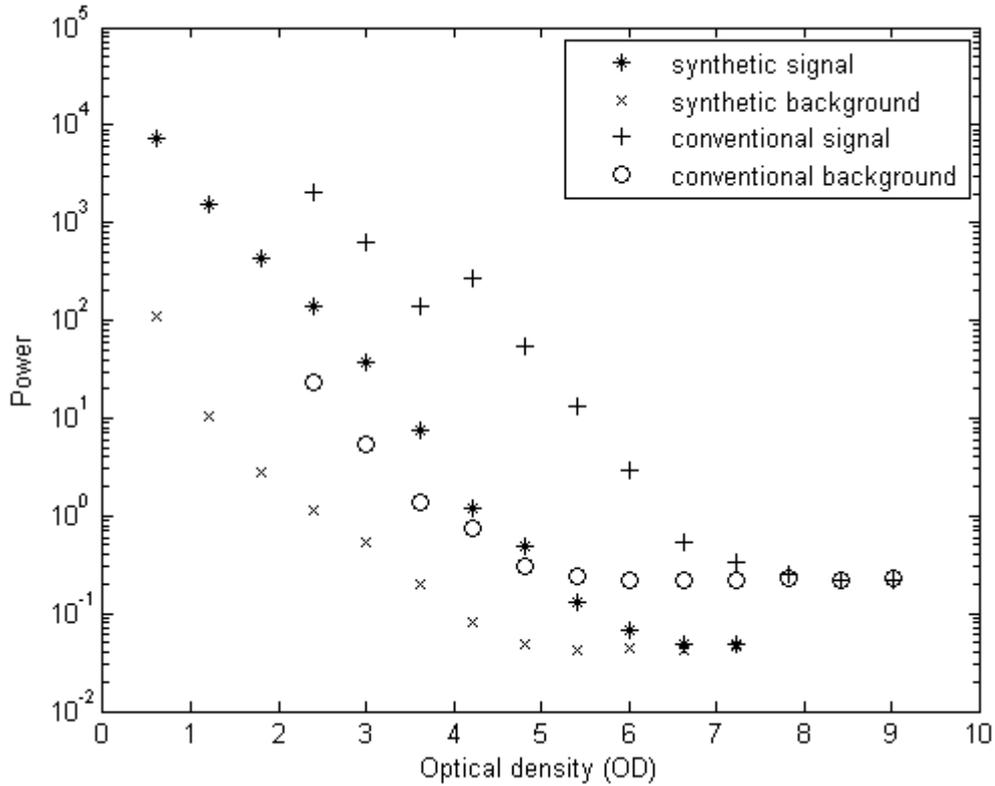

**Figure 8:** Power measurements from the target and from the background versus optical density. The setup is the same as in Figure 6 but with an attenuator before the acousto-optic deflectors. The setups are quantum noise limited. Again the when the round trip density is increased by one, le power is divided by ten which is coherent with Beer-Lambert law.

Both conventional and SA curves decrease with the optical attenuation with Beer-Lambert law, which corresponds to the attenuation of the power linked to parasitic reflections. The difference between the two shot noise limits for conventional and SA imaging comes from the adapted filtering: in SA modality, a big part of the noise is filtered whereas in conventional imaging, the image is free of any numerical treatment. This figure indicates that the laser quantum shot noise is about 3.5 orders of magnitude below our limitation for the synthetic LOFI microscope and 4.5 orders of magnitude for the conventional microscope. This shows that there is a big potential of improvement for both microscopes. Moreover, knowing that the LOFI setup is shot noise limited, noise corresponds to one photon reinjected during integration time (here T = 100 μs), the figure 8 gives an idea of the number of photons reinjected from both signals and backgrounds. More precisely, there are approximately $3.10^3$ parasitic photons in the synthetic setup and $3.10^4$ for conventional microscope.

## 6) Conclusions and outlook

In this paper we compared the performances of two LOFI-based microscopes: a conventional one, and another based on Synthetic Aperture and numerical post-treatment. In particular we have shown that the resolution is strictly equivalent in both cases and degrades linearly with distance. We then theoretically and experimentally proved that the sensitivity of both microscopes degrades with the power four of the working distance. At equal resolutions, direct imaging microscope is more sensitive (factor 2800 with the parameters used in the paper) and more photons are collected in this configuration. Images have been realised through milk diluted in water and have shown to be limited by parasitic reflections on elements between the mirrors and the target. Finally, the measurements have shown that with our parameters the SNR (ratio of signal photons to parasitic reflected photons) performances of conventional LOFI is 68 times higher that SA microscope configuration. So benefits of the holographic microscope (3D imaging in a single x-y acquisition) are obtained at the expense of a degradation of the photometric performances.

In the future, several improvements are going to be explored. First it has been shown that a translational scanning rather than the current galvanometric mirror scanning leads to a conservation of the resolution whatever is the working distance L. But this configuration has not been experimented yet because it implies mechanical movements which are both slow and cause of important vibrations.

Another important improvement is suggested by the last study of this paper showing that our current background limitation for the holographic microscope is 3000 times above the shot noise. The parasitic reflections could be suppressed totally if only photons that are reflected by the target are shifted by the good frequency [29,30]. This improvement could lead us to a shot noise limitation and so to an SNR improvement of three orders of magnitude.

To conclude, the phase of the synthetic signal could be also used to realise holographic resolved profilometry or displacement maps. It is already used in SAR for geosciences applications [31]. It still has to be applied to SAL in the next future.

REFERENCES

1. E. Lacot, O. Jacquin, G. Roussely, O. Hugon, and H. Guillet de Chatellus, "Comparative study of autodyne and heterodyne laser interferometry for imaging," J. Opt. Soc. Am. A **27**, 2450-2458 (2010).

2. O. Jacquin, E. Lacot, W. Glastre, O. Hugon, and H. Guillet de Chatellus, "Experimental comparison of autodyne and heterodyne laser interferometry using Nd:YVO4 microchip laser," J. Opt. Soc. Am. A **28**, 1741-1746 (2011).

3. S. M. Popoff, G. Lerosey, R. Carminati, M. Fink, A. C. Boccara, and S. Gigan, "Measuring the Transmission Matrix in Optics : An Approach to the Study and Control of Light Propagation in Disordered Media," Phys. Rev. Lett. **104**, 100601 (2010).

4. M. Pernot, J-F. Aubry, M. Tanter, A-L. Boch, F. Marquet, M. Kujas, D. Seilhean, and M. Fink, "In vivo transcranial brain surgery with an ultrasonic time reversal mirror," J. Neurosurg. **106**, 1061-1066 (2007).

5. A. Dubois, C. Boccara, "Full-filed OCT," M S-Medecine Sciences. **22**, 859-864 (2006).

6. M. Minsky, "Memoir on inventing the confocal scanning microscope," Scanning. **10**, 128-138 (1988).

7. I. M. Vellekoop, C. M. Aegerter, "Scattered light fluorescence microscopy : imaging through turbid layers," Opt. Letter. **35**, 1245-1247 (2010).

8. P. Pantazis, J. Maloney, D. Wu, S. E. Fraser, "Second harmonic generating (SHG) nanoprobes for in vivo imaging," PNASUSA. **107**, 14535-14540 (2010).

9. S. Vertu, J. Flugge, J.J. Delaunay, and O. Haeberle, "Improved and isotropic resolution in tomographic driffractive microscopy combining sample and illumination rotation," CENTRAL EUROPEAN JOURNAL OF PHYSICS. **44**, 969-974 (2011).

10. K. Otsuka, "Self-Mixing Thin-Slice Solid-State Laser Metrology," Sensors. **11**, 2195-2245 (2011).

11. A. Witomski, E. Lacot, O. Hugon, and O. Jacquin, "Two dimensional synthetic aperture laser optical feedback imaging using galvanometric scanning," Appl. Opt. **47**, 860–869 (2008).

12. E. Lacot, R. Day, and F. Stoeckel, "Laser optical feedback tomography", Opt. Lett. **24**, 744–746 (1999).

13. E. Lacot, R. Day, and F. Stoeckel, "Coherent laser detection by frequency-shifted optical feedback," Phys. Rev. A **64**, 043815 (2001).

14. R. Day, E. Lacot, F. Stoeckel, and B. Berge, "Three-dimensional sensing based on a dynamically focused laser optical feedback imaging technique," Appl. Opt. **40**, 1921-1924 (2001).

15. O. Hugon, E. Lacot, and F. Stoeckel, "Submicrometric Displacement and Vibration Measurement Using Optical Feedback in a Fiber Laser," Fib. Integr. Opt. **22**, 283-288 (2003).

16. O. Hugon, I. A. Paun, C. Ricard, B. van der Sanden, E. Lacot, O. Jacquin, and A. Witomski, "Cell imaging by coherent backscattering microscopy using frequency-shifted optical feedback in a microchip laser," Ultramicroscopy. **108**, 523-528 (2008).

17. O. Hugon, F. Joud, E. Lacot, O. Jacquin, and H. Guillet de Chatellus, "Coherent microscopy by laser optical feedback imaging (LOFI) technique," Ultramicroscopy. **111**, 1557-1563 (2011).

18. V. Muzet, E. Lacot, O. Hugon, and Y. Guillard, "Experimental comparison of shearography and laser optical feedback imaging for crack detection in concrete structures", Proc. SPIE **5856**, 793-799 (2005).

19. K. Otsuka, T. Ohtomo, H. Makino, S. Sudo, and J. Ko, "Net motion of an ensemble of many Brownian particles captured with a self-mixing laser," Appl. Phys. Letters. **94**, 241117 (2009).

20. S. Sudo, T. Ohtomo, Y. Takahashi, T. Oishi, and K. Otsuka, "Determination of velocity of self-mobile phytoplankton using a self-mixing thin-slice solid-state laser," Appl. Opt. **48**, 4049-4055 (2009).

21. J. C. Curlander, and R. N. McDonough, *Synthetic Aperture Radar: Systems and Signal Processing*, (Wiley, 1991).

22. A. Ja. Pasmurov, and J. S. Zimoview, *Radar Imaging and Holography*, (Institution of Electrical Engineers, 2005).

23. C.C. Aleksoff, J.S. Accetta, L.M.Peterson, A.M.Tai, A.Klossler, K. S. Schroeder, R. M. Majwski, J. O. Abshier, and M. Fee, "Synthetic aperture imaging with a pulsed CO2 laser," Proc. SPIE **783**, 29–40 (1987).

24. S. Markus, B. D. Colella, and T. J. Green, "Solid-state laser synthetic aperture radar," Appl. Opt. **33**, 960 – 964 (1994).

25. A. Witomski, E. Lacot, O. Hugon, and O. Jacquin, " Synthetic aperture laser optical feedback imaging using galvanometric scanning," Opt. Lett. **31**, 3031 – 3033 (2006).

26. F. Dubois, C. Schockaert, N. Callens, C. Yourassowsky, "Focus plane detection criteria in digital holography microscopy by amplitude analysis," Opt. Express, **14**, 5895–5908 (2006).

27. M.I. Kolobov, L. Davidovich, E. Giacobino, and C. Fabre, "Role of pumping statistics and dynamics of atomic polarization in quantum fluctuations of laser sources," Phys. Rev. A **47**, 1431-1446 (1993).

28. A. Bramati, J.P. Hermier, V. Jost, E. Giacobino, L. Fulbert, E. Molva, and J.J. Aubert, "Effects of pump fluctuations on intensity noise of Nd:YVO4 microchip lasers," Eur. Phys. J. D. **6**, 513-521 (1999).

29. O. Jacquin, E. Lacot, C. Felix, and O. Hugon, "Laser optical feedback imaging insensitive to parasitic optical feedback," Appl. Opt. **46**, 6779-6782 (2007).

30. O. Jacquin, S. Heidmann, E. Lacot, and O. Hugon, "Self-aligned setup for laser optical feedback imaging insensitive to parasitic optical feedback, " Applied Optics, **48**, 64-68 (2009).

31. G. Liu, J. Li, Z. Xu, J. Wu, Q. Chen, H. Zhang, R. Zhang, H. Jia, and X. Luo, "Surface deformation associated with the 2008 Ms8.0 Wenchuan earthquake from ALOS L-band SAR interferometry," International Journal of Applied Earth Observation and Geoinformation. **12**, 496-505 (2010).